\documentclass[twocolumn,usenatbib,useAMS]{mn2e}
\usepackage{natbib}
\usepackage{amsbsy}
\usepackage{amssymb}
\usepackage{amsmath}
\usepackage{graphicx}
\usepackage{xcolor}

\begin{document}

\title[Extension of the Vortex Creep Model] {Peculiar Glitch of PSR J1119-6127 and Extension of the Vortex Creep Model}

\author[Akbal, G\"ugercino\u{g}lu, \c{S}a\c{s}maz Mu\c{s}, Alpar]{ O. Akbal$^1$\thanks{E-mail:
onurakbal@sabanciuniv.edu}, E. G\"{u}gercino\u{g}lu$^{2}$\thanks{E-mail:
egugercinoglu@gmail.com}, S. \c{S}a\c{s}maz Mu\c{s}$^1$\thanks{E-mail:
sinemsm@sabanciuniv.edu}, M.A. Alpar $^{1}$\thanks{E-mail:
alpar@sabanciuniv.edu} \\
$^1$Sabanc{\i} University, Faculty of Engineering and Natural Sciences, Orhanl{\i}, 34956 Istanbul, Turkey \\
  $^2$Istanbul University, Faculty of Science, Department of Astronomy and Space Sciences, Beyaz{\i}t, 34119, Istanbul, Turkey}

\maketitle

\begin{abstract}
Glitches are sudden changes in rotation frequency and spin-down rate,
observed from pulsars of all ages. Standard glitches are characterized
by a positive step in angular velocity ($\Delta\Omega$ $ > $ $0$) and
a negative step in the spin-down rate ($\Delta \dot \Omega$ $ < $ $0$)
of the pulsar. There are no glitch-associated changes in the electromagnetic
signature of rotation-powered pulsars in all cases so far. For the first time,
in the last glitch of PSR J1119-6127, there is clear evidence for changing
emission properties coincident with the glitch. This glitch is also unusual in
its signature. Further, the absolute value of the spin-down rate actually decreases in the long term. 
This is in contrast to usual glitch behaviour. In this paper we extend the vortex creep
model in order to take into account these peculiarities. We propose
that a starquake with crustal plate movement towards the rotational
poles of the star induces inward vortex motion which causes the unusual glitch signature. 
The component of the magnetic field perpendicular to the rotation axis will decrease, 
giving rise to a permanent change in the pulsar external torque.  
\end{abstract}


\section{Introduction}
\label{sec:intro}

The 2007 glitch of PSR J1119-6127 is unusual and interesting as the first case with
clear indications of changing pulsar emission properties coincident with the glitch.
The event is also unusual in its long term signature of {\em decreased} spin-down rate. These signatures require
an extension of the vortex creep model which has become the standard model for
evaluating glitches and post-glitch response. The extension of the model must also
make allowance for changes in the pulsar torque suggested by the glitch related
changes in emission properties.

Glitches are sudden increases in the rotation rate of pulsars followed by relaxation
towards the pre-glitch state. The fractional change of the angular velocity, $\Delta\Omega/\Omega$, in a glitch is in the range $\sim 10^{-10} - 10^{-5}$.
Glitches are usually accompanied by jumps in the spin-down rate,
$\Delta \dot \Omega /\dot \Omega$, in the range $\sim 10^{-4} - 10^{-2}$.
To date, about 400 glitches have been observed in more than a hundred pulsars
\citep{espinoza11,yu13}. Since the earliest glitch observations,  the sudden changes
in rotation frequency and spin-down rate were seen to relax back
towards the pre-glitch values on timescales of days to years. This is interpreted as a
signature of superfluid interior components of the neutron star \citep{baym69}, as a
star composed of normal matter would relax much faster.

Several models have been proposed to explain glitches and post-glitch relaxation. In the
early starquake model \citep{ruderman69}, the solid crust of the neutron star occasionally cracks under
stresses induced by the ongoing spin-down of the star, thereby readjusting to a less oblate shape closer to the equilibrium shape that a fluid star would follow while
spinning down. By conservation of angular momentum, the reduction in moment of inertia
of the crust is accompanied by an increase in its angular velocity.
Glitches in the Crab pulsar \citep{wong01} and PSR J0537-6910 \citep{middleditch06}
can be explained by this model. However,  starquakes cannot explain large glitches that
repeat every few years, as exhibited by the Vela pulsar \citep{baym71}. The required rate
of dissipation of elastic energy stored in the solid crust would produce an X-ray luminosity
enhancement which is not observed \citep{gurkan00}.

The standard model for the pulsar glitches is the vortex pinning$-$unpinning (vortex creep) model  based on the dynamics of the neutron star's superfluid interior \citep{anderson75, alpar84}. This model invokes the minimal storage and dissipation of energy
for a star with angular momentum.
The expected energy dissipation in a large glitch, at the expense
of the rotational kinetic energies of the two components, does not violate any observational
upper bounds. Models based on pinned superfluid components can explain the various modes of glitch and post-glitch behaviour \citep{haskell12,haskell14}.

Radio pulsar glitches observed up to the 2007 glitch of PSR
J1119-6127 \citep{weltevrede11} showed no glitch correlated changes
in the electromagnetic signatures, like pulse shape, emission
pattern, spectrum and polarization. Previous applications of
the vortex creep and starquake models assumed that there were no changes
in the pulsar torque at the time of the glitch. Glitches
and post-glitch response were explained entirely in terms of the
internal structure and dynamics of the neutron star. The 2007 glitch of PSR J1119-6127 
shows clear evidence for changing emission properties induced by the
glitch, switching on intermittent pulses
\citep[see, e.g.,][]{kramer06} and also showing
rotating radio transient (RRAT) behaviour
\citep[see, e.g.,][]{keane11}. Interestingly, this glitch also
displayed $\Delta \dot \Omega > 0$ after transients have
decayed, in contrast to the signatures of ``standard" glitches which are
characterized by a negative step in spin-down rate ($\Delta \dot \Omega < 0$). The 
high magnetic field radio pulsar PSR J1846-0258 had comparable glitch-induced
emission changes \citep{livingstone10}. The radio pulsar PSR J0742-2822 showed a suggestive 
connection between changing radio emission and
pulse shape features and glitch activity, however there is currently little direct 
evidence to establish a robust link between them due to absence of enough data following 
the glitch date \citep{keith13}. The RRAT J1819-1458 was also
reported to have an increase in its activity associated with a
glitch \citep{lyne09}. 

In this paper we analyze the 2007 glitch of PSR J1119-6127 and extend the vortex creep model to include
the possibility of a sudden change in the pulsar torque associated with
the glitch, as suggested by the changing emission properties, and to address the
atypical glitch signature.  In \S\ref{sec:glitch} we summarize the unique properties
of the 2007 glitch of PSR J1119-6127. In \S\ref{sec:overview} we review the vortex
creep model, while in \S\ref{sec:extension} we develop the model to include the
unusual signatures in the spin frequency and spin-down rate, and allow for glitch
associated changes in the pulsar torque.
We apply our extended model to the peculiar glitch of PSR J1119-6127 in \S\ref{sec:fit}.  We discuss
our results in \S\ref{sec:discussion}.

\section{The Peculiar Glitch of PSR J1119-6127}
\label{sec:glitch}

PSR J1119-6127 is a young pulsar with a period $P = 0.41$ s and a period derivative
$\dot{P} = 4\times10^{-12} $ Hz s$^{-1}$ discovered by \citet{camilo00}. It has
a characteristic age $\tau_{c} \equiv P / ( 2 \dot{P}) \cong $ 1625 years, and a high
surface dipole magnetic field $B \sim 8.2 \times 10^{13}$ G (at the poles). 
This pulsar has exhibited three glitches \citep{camilo00, weltevrede11}.
The third glitch, which occurred in 2007, was quite unusual in a number of ways
\citep{weltevrede11}:

\begin{enumerate}

\item For a while after the initial exponential relaxation is completed, the pulsar is found to be rotating
with a smaller angular velocity as compared to the pre-glitch value, $\Delta\Omega (t) < 0$. 
In the latest data $\Delta\Omega (t) > 0$, and may be settling at a positive value \citep{antonopoulou14}. 

\item In the long term, the  pulsar slows down at a lower rate; the absolute value of the
spin-down rate is less (the frequency derivative is greater)
than its pre-glitch value, $\Delta \dot \Omega$ $ > $ 0.

\item While the fractional changes in angular velocity are small, of the order of 10$^{-9}$, for the glitches of the Crab pulsar, Vela and  older pulsars undergo large glitches of size $\Delta\Omega/\Omega \sim$10$^{-6}$ as well as smaller ``Crab-like"
events. The 2007 glitch of  PSR J1119-6127 is a ``Vela-like" giant glitch 
from a {\em young} pulsar comparable to the Crab pulsar in characteristic age.

\item The radio emission properties of PSR J1119-6127 displayed changes associated
with the glitch. The pulsar switched on intermittent pulses and also showed RRAT behaviour which
seems to have emerged with the glitch. This anomalous emission behaviour of
PSR J1119-6127 was observed for about three months following the 2007 glitch.

\end{enumerate}

The much smaller second glitch which occurred in 2004 may have had similar signatures in the long term post-glitch frequency and frequency derivative remnants \citep{antonopoulou14}. The data is sparse, and post-glitch evolution may have been interrupted by the arrival of the 2007 glitch. Furthermore, no glitch associated changes in emission properties were observed for the 2004 glitch. Here we address only the 2007 glitch.


\section{Overview of the Vortex Creep Model}
\label{sec:overview}


The vortex creep model \citep{alpar84,alpar89} attempts to explain the processes which
cause both the glitches and the post-glitch relaxation in terms of a
number of distinct superfluid regions in the inner crust. The superfluid core of the neutron star is coupled to the external torque 
on very short timescales, via electron scattering off magnetized vortices \citep{alpar84b}. The core superfluid therefore behaves as part of the effective normal matter crust. Hence the superfluid component relevant for glitch and postglitch dynamics is the crust superfluid. A description of the core superfluid blue as well as the crustal superfluid in terms of mutual friction forces acting upon vortex lines is given by \citet{andersson06}.

The dynamics of the crust superfluid is constrained by the pinning of the quantized vortex lines to nuclei, interstitial positions and possibly other structures in the crust lattice \citep{alpar77,link91, mochizuki99, avogadro08, pizzochero11, haskell12, seveso14}. When vortices pin to nuclei, they move with the crust's velocity.
A lag  $\omega=\Omega_{s} - \Omega_{c}$ builds up between the superfluid and the crustal
angular velocities $\Omega_{s}$ and $\Omega_{c}$ as the crust spins down under the
external pulsar torque. This lag is sustained by the pinning forces acting upon the vortex line. In the case of rotational (cylindrical) symmetry, the magnitude of the required pinning force (per unit length) is
$f=r\rho_{s}\kappa\omega=r\rho_{s}\kappa(\Omega_{s}-\Omega_{c})$, where
$r$ is the distance from rotation axis, $\rho_{s}$ is superfluid density, $\kappa$ is the quantum of vorticity. 
The critical (maximum) lag, $\omega_{cr}$, determined by the maximum available pinning force, is given by $\omega_{cr}=
E_{p}/rb\rho_{s}\kappa\xi$. Here $E_{p}$ is pinning energy, $\xi$ is the vortex core
radius and $b$ is the distance between successive pinning sites along the vortex line. If local fluctuations 
in vortex density and superfluid velocity raise $\omega$ above $\omega_{cr}$, there will be sudden
unpinning and outward motion which can lead to an avalanche of vortex discharge \citep{anderson75}.
By conservation of angular momentum this leads to speeding up of the crust, $\Delta\Omega_{c} >  0$, observed
as a glitch. The possibility of such
vortex unpinning avalanches taking place spontaneously was confirmed by computer simulations \citep{melatos09, warszawski11, warszawski12}. 

Apart from the discontinuous angular momentum imparted to the crust by sudden vortex
unpinning at glitches, the superfluid also spins down continuously
between glitches by outward flow of vortices. The crustal neutron superfluid follows the spin-down of the crust by means of thermally activated outward creep of vortex lines against the pinning
energy barriers \citep{alpar84, alpar89}.

In terms
of a simple two component model, involving the crust and the superfluid component,
the observed spin-down of a neutron star's crust satisfies the equation,
\begin{equation}
I_{c}\dot\Omega _{c} = N_{ext} + N_{int} = N_{ext} - I_{s}\dot\Omega _{s},
\end{equation}
where $N_{ext} =I\dot\Omega _{\infty}$ is the external torque on the neutron star which
tries to slow down the crust, and $N_{int}$ is the internal torque arising from the coupling of the
superfluid to the crust by vortex creep and tends to speed up the crust.
$I_{c}$ is the moment of inertia of the effective crust (including the superfluid core of the neutron star), $I_{s}$ is the moment of inertia of the pinned superfluid, while their spin-down
rates are $\dot\Omega _{c}$ and $\dot\Omega _{s}$, respectively. The spin-down rate $ \dot\Omega _{s} $ of the superfluid is determined by vortex creep \citep{alpar84}. The system reaches a steady state when both the superfluid and the crust spin-down at the same rate $\dot\Omega _{\infty} \equiv N_{ext}/(I_{s}+I_{c})$, sustained at the steady state lag $ \omega _{\infty} $.  

Glitches set the system off from steady state. Post-glitch relaxation is due to the
recovery of vortex creep, as the lag $\omega$ inevitably builds back towards 
steady state due to the ongoing spin-down of the crust under the external pulsar
torque. The internal torque is so sensitively dependent on the pinning energy
$E_{p}$ and the crustal temperature $T$ that we expect vortex lines
in the different regions of the superfluid to respond differently.
Depending on the temperature and the local pinning parameters in relation to the external torque,
vortex creep can have a linear or nonlinear dependence on the lag
\citep{alpar89}\footnote{The claim that the linear regime of vortex creep is never realized for realistic pinning parameters \citep{link14} depends on the velocity of unpinned vortices, relying on the assumption that they move with the global averaged superfluid velocity with drag forces, and are not affected by the contributions of interactions with the adjacent pinning sites to the local superfluid velocity. This issue will be addressed in a separate work.}. In the linear regime, the response is linear in the glitch-induced perturbation to the lag $ \omega $ and gives simple exponential relaxation. The relaxation time $\tau_{l}$ is very sensitively dependent on $E_p/kT$,
with $\tau_{l} \propto exp(E_p/kT)$. The steady state lag $\omega_{\infty} = \vert\dot{\Omega}\vert_{\infty}\tau_{l}$ is always much less than $\omega_{cr}$ in this regime. From glitch observations, up to four exponential relaxation terms are seen
from a particular pulsar \citep{dodson07}.

In the opposite regime we have a very nonlinear response to perturbations. The response of a nonlinear creep region $k$ to the glitch
will be \citep{alpar84},
\begin{equation}
\Delta{\dot\Omega_{c,k}}=-\frac{I_{k}}{I}\vert\dot{\Omega}\vert_{\infty}
\left[1-\frac{1}{1+({\rm e}^{t_{0,k}/\tau_{nl}}-1){\rm
e}^{-t/\tau_{nl}}}\right].
\end{equation}
At the time of glitch, creep in those regions which show nonlinear
response can stop temporarily. These regions decouple from rest of
the star, so that external torque acts on less moment of inertia.
Creep restarts after a waiting time of $t_{0} =
\delta\omega/\vert\dot{\Omega}\vert_{\infty}$, since the external torque
restores the glitch induced decrease in angular velocity lag. The relaxation time is
\begin{equation}
\tau_{nl}=\frac{kT}{E_{p}}\frac{\omega_{cr}}{|\dot\Omega|_{\infty}}. \nonumber
\end{equation}

In those superfluid regions through which the avalanche of vortices unpinned at the glitch pass, moving rapidly in the radially outward direction, the ensuing reduction $\delta\Omega_s$ in the superfluid rotation rate determines the offset $\delta\omega = \delta\Omega_s + \Delta\Omega_c$ in the lag, as $\delta\Omega_s \gg \Delta\Omega_c$. This results in the response given in Eq. (2), characterized by the waiting time $t_{0} \cong \delta\Omega_s/\vert\dot{\Omega}\vert_{\infty}> \tau_{nl}$. There can also be nonlinear creep regions through which no unpinned vortices pass at the glitch, so that $\delta\omega = \Delta\Omega_c$. In this case $t_0 = \Delta\Omega_c/\vert\dot{\Omega}\vert_{\infty}$ can be much shorter than $\tau_{nl}$, and the contribution of such a nonlinear creep region reduces to simple exponential relaxation \citep{erbil14},
\begin{equation}
\Delta{\dot\Omega_{c,k}}\cong -\frac{I_{k}}{I}\frac{\Delta\Omega_c}{\tau_{nl}}
{\rm e}^{-t/\tau_{nl}}
\end{equation}
like in the case of linear creep regions, but with the nonlinear creep relaxation time $\tau_{nl}$.

If we integrate Eq. (2) with the assumption that the post-glitch superfluid angular velocity
decreases linearly in $r$ over the region, corresponding to uniform density of unpinning vortices, one obtains \citep{alpar84} 

\begin{equation}
\noindent
\frac{\Delta\dot{\Omega}_{c}(t)}{\dot{\Omega}_{c}}=\frac{I_{A}}{I}
\left\lbrace1-\frac{1-(\tau_{nl}/t_{0})\ln\left[1+(e^{t_{0}/\tau_{nl}}-1)e^{-\frac{t}{\tau_{nl}}}\right]}{1-e^{-\frac{t}{\tau_{nl}}}}\right\rbrace.
\end{equation}
In the limit $t_{0} \gg \tau_{nl}$ this reduces to recovery with a constant $\ddot{\Omega}_{c}$
\begin{equation}
\frac{\Delta \dot{\Omega}_{c}(t)}{\dot{\Omega}_{c}} = \frac{I_{A}}{I} \left( 1 - \frac{t}{t_{0}} \right),
\end{equation}
as observed in the Vela pulsar \citep{alpar93} and in most Vela-like giant glitches in older pulsars
\citep{yu13}. In the above equations $t_{0}$ is the maximum waiting time, $I_{A}$ is the
moment of inertia of the vortex creep region A where unpinning of the vortices has taken place during the glitch. Vortices unpinned in regions A pass through regions B with moment of inertia $I_{B}$ before repinning in another creep region A. Regions B do not participate in spin-down by creep, as they do not sustain pinned vortices. Regions B contribute to the angular momentum transfer only at glitches, when an avalanche of unpinned vortices moves through them. These regions A and B determine the glitch, interglitch and long term behaviour of pulsars \citep{alpar93,alpar96}.

After the exponential transients are removed, observable
variables associated with glitches are related to the model parameters by
the following simple three equations \citep{alpar06}:
\begin{equation}
I_{c}\Delta \Omega _{c} = (I_{A}/2 + I_{B}) \delta \Omega_{s}.
\end{equation}
\begin{equation}
\frac{\Delta \dot \Omega _{c}}{\dot \Omega _{c}}=\frac{I_{A}}{I}.
\end{equation}
\begin{equation}
\ddot{\Omega} _{c} = \frac{I_{A}}{I} \frac{{\dot\Omega}_{\infty}^2}{\delta\Omega_{s}} .
\end{equation}
Eq. (6) simply states
angular momentum conservation and gives the glitch magnitude. This is proportional to the number of vortices which participated in the glitch event. For a uniform array of vortices the number of unpinned vortices moving outward through radius $r$ is related to the change
in angular velocity of the superfluid at $r$,
\begin{equation}
\delta N = 2 \pi r^{2}\delta\Omega_{s} / \kappa \cong 2 \pi R^{2}\delta\Omega_{s} / \kappa,
\end{equation}
since $r \cong R$, the radius of the star, in the crust superfluid. The angular momentum transfer depends on $\delta \Omega_{s}$ and the moment of inertia of the regions that vortices pass through, $I_{A}$ and $I_{B}$. Eq. (7) is about the torques acting on the pulsar. Before the glitch, in steady state, the crust superfluid and the rest of the star spin down at the same rate. When a glitch occurs, some part of the crustal superfluid decouples
from the external torque leading to a jump in spin-down rate. Solving these equations for the three unknowns, $I_{A}$, $I_{B}$, and $\delta\Omega_{s}$, one can obtain model parameters uniquely without making any further assumptions.

\section{Extension of the Vortex Creep Model}
\label{sec:extension}

In the standard vortex unpinning-creep model only the outward motion
of vortices is considered. This gives a negative post-glitch offset (an increase in the absolute value) of the spin-down
rate from its pre-glitch value, $\Delta\dot\Omega<0 $. The spin-down rate relaxes back to the pre-glitch
value ($\Delta\dot\Omega\rightarrow 0$) for all modes of vortex creep which supply the internal torques from
the superfluid acting on the normal matter crust. Thus, (i) glitches with
the ``wrong" sign in frequency and spin-down rate require
inward vortex motion at the glitch; and (ii) long term (persistent) shifts
in the spin-down rate require either a structural change
in the neutron star crust, as proposed for the Crab pulsar \citep{alpar96},
or a glitch associated shift in the external torque \citep{link92}.

Occasional inward fluctuations of vortices, facing an extra potential
barrier, is a low probability component of the creep process. Therefore, bulk spontaneous
inward motion of an avalanche of unpinned vortices is thermodynamically
impossible in an isolated superfluid. Large numbers of vortices could be transported inward only if the
glitch were induced by an agent external to the superfluid, like a starquake.

Inward vortex motion will increase the superfluid velocity by some $\delta\Omega'_s$ in regions of superfluid through which vortices have moved inward. Its effect can be investigated by changing
$t_0$ with $- t'_0$, where $t'_0 \cong \delta\Omega_s'/\vert\dot{\Omega}\vert_{\infty}$. With this we obtain:
\begin{equation}
\Delta{\dot\Omega_c}=-\frac{I_{A'}}{I}\vert\dot{\Omega}\vert_{\infty}
\left[1-\frac{1}{1+({\rm e}^{-t'_0/\tau'_{nl}}-1){\rm e}^{-t/\tau'_{nl}}}\right],
\end{equation}
where the primes indicate parameters associated with inward vortex motion.
This equation describes the response to inward motion of unpinned vortices.
When vortices travel inward, superfluid rotates faster. The lag $\omega$
thereby increases from its steady state value, and creep will be more
efficient than in steady state, with an enhanced vortex current in the
radially outward direction. If we integrate  Eq. (10) over a nonlinear creep
region throughout which a uniform average density of vortex lines unpinned,
or repinned, we obtain:
\begin{equation}
\frac{\Delta\dot{\Omega}_{c}(t)}{\dot{\Omega}_{c}}=
\frac{I_{A'}}{I}\left\lbrace1-
\frac{1+(\tau'_{nl}/t'_{0})\ln\left[1+(e^{-t'_{0}/\tau'_{nl}}-1)e^{-\frac{t}{\tau'_{nl}}}
\right]}{1-e^{-\frac{t}{\tau'_{nl}}}}\right\rbrace .
\end{equation}
The internal torque contribution given in Eqs. (10) and (11) leads to an initial
positive contribution to $ \Delta\dot{\Omega}_{c} $, which asymptotically
decays to zero. Unlike the nonlinear creep response to glitch associated {\em outward} vortex motion, as given in Eq. (2), the nonlinear creep response to {\em inward} 
vortex motion, {\em does not have a waiting time}. Instead Eqs. (10) and (11) display quasi-exponential relaxation. {\em A constant second derivative} $\ddot{\Omega} _{c}$ {\em is not obtained} from Eq. (11) when $t'_{0} \gg \tau'_{nl}$ or in any other limit. As the integrated response in Eq.(11) is very similar, Eq. (10) is adequate to describe the spindown rate when vortices have moved inward.

Allowing for the starquake induced inward vortex motion at the glitch, in
addition to the natural outward motion of many unpinned vortices, we get
the following equation instead of Eq. (6),
\begin{equation}
I_{c}\Delta \Omega _{c}(0) = (I_{A}f + I_{B}) \delta
\Omega_{s}-(I_{A'}f + I_{B'}) \delta \Omega'_{s}.
\end{equation}
where $f=1/2$ is for the integrated response, Eqs.(4), (5) and (11), and $f=1$
for the simpler response, Eqs.(2) and (10). The first term on the right hand side is
the angular momentum transfer due to outward moving vortices, while the second term is the contribution of inward moving
vortices. The physical meanings of $I_{A'}$ and $I_{B'}$ are similar to
their non-primed counterparts. A plate of the crustal
solid that moves inward in a quake could carry vortices with it, in the inward,
$-r$, direction. Nonlinear creep
regions with moment of inertia $I_{A'}$, and vortex free regions with moment of inertia $I_{B'}$ are at radial positions between the original and the new positions of the plate, 
and therefore experience a sudden increase $\delta\Omega_s '> 0$. As creep relaxes back to steady state, the net angular momentum transfer from the regions $A$ and $A'$ is zero, while the regions $B$ and $B'$ transport 
angular momentum only at glitches and will contribute a remnant frequency offset $\Delta\Omega_p$:
\begin{equation}
I_{c}\Delta \Omega _{p} = I_{B} \delta
\Omega_{s}- I_{B'} \delta \Omega'_{s}.
\end{equation}

Extending Eq. (7) to
describe the net glitch in the spin-down rate with the terms of
opposite signs describing the response of creep to outward and inward
vortex motion, we obtain
\begin{equation}
\frac{\Delta \dot \Omega_{c}}{\dot \Omega_{c}}=\frac{I_{A}}{I}-\frac{I_{A'}}{I}.
\end{equation}

For PSR J1119-6127 the post-glitch $\Delta \dot \Omega > 0$ persists for $\sim 2500$ days,
as far as the pulsar has been observed since the glitch \citep{antonopoulou14}. Here we pursue the assumption that $\Delta \dot{\Omega}_{c} > 0$ is permanent; that it will not decay on long timescales in the future. This is a viable assumption with the present data, as discussed in the next section. With this assumption the permanent shift $\Delta \dot \Omega_{p}$ could be due to
a structural change in the star, as postulated for the persistent shifts in spin-down
rate observed to accompany the Crab pulsar glitches \citep{alpar96}, or, alternatively,
due to a glitch associated permanent change in the external torque. Unlike the Crab
pulsar, the 2007 glitch of PSR J1119-6127 has strong indications that actually the external
torque has changed, since the pulsar has switched to intermittent and RRAT behaviour
with the glitch. It is likely that structural changes experienced by PSR J1119-6127   
lead to a permanent change in the external torque.

\section{Model Fits}
\label{sec:fit}

We apply a model which is an extension of
earlier applications of the vortex creep model to the Vela
\citep{alpar93} and Crab \citep{alpar96} pulsars' glitches. We take
one nonlinear creep region with relaxation time $ \tau_2$ corresponding
to the outward motion of the glitches. The new component in the
extended model is the inclusion of inward moving vortices in the
glitch, which move through a nonlinear creep region with relaxation
time $ \tau_1$ (cf. Eq. (10)). We also
employ a region in which relaxation occurs exponentially with a timescale $\tau_3$, discussed below.
Finally, we include a possible external torque change as a constant offset
to the spin-down. We tried model fits with the
integrated response, Eqs.(4) and (11) and with the simple response Eq.(2) and (10).
As the residuals are comparable, we choose to employ the simple model.

The expression used for the fit including our extended formula is:
\begin{align}
\Delta\dot{\Omega}_{c}(t)=
&-a_{1}\left[1-\frac{1}{1+\alpha_{1}e^{-(t+\Delta)/\tau_{1}}}\right]
\nonumber\\
&-a_{2}\left[1-\frac{1}{1+\alpha_{2}e^{-(t+\Delta)/\tau_{2}}}\right]
-a_{3}e^{-(t+\Delta)/\tau_{3}}+b.
\end{align}
The various parameters are defined by: $
a_{1}=\frac{I_{A'}}{I}|\dot{\Omega}|_{\infty} $, $
a_{2}=\frac{I_{A}}{I}|\dot{\Omega}|_{\infty} $, $
\alpha_{1}=(e^{-t_{0}'/\tau_{1}}-1) $, $
\alpha_{2}=(e^{t_{0}/\tau_{2}}-1) $, $
a_3=\frac{I_{3}}{I}\frac{\delta\omega}{\tau_{3}}$, and $ b=(\Delta
N_{ext}/N)\dot{\Omega}_{\infty} $, and $t$ is the time since the
first post-glitch observation, with the time lag $\Delta$ between the
actual glitch date and the first post-glitch observation. We
have 9 free parameters. Parameters with the subscript ``1" denote the
contribution from the response of vortex creep to glitch associated
inward vortex motion, while those with subscript ``2" and ``3" are associated
with creep response to glitch associated outward vortex motion.

The exponential relaxation term with amplitude $ a_{3} $
might describe the response of either an intrinsically linear creep region,
or a nonlinear creep region where there was no vortex motion at the glitch, so that the
angular velocity of the superfluid remains unchanged and the glitch
induced perturbation to the angular velocity lag is simply
$\delta\omega = \Delta\Omega_{c}$ \citep{erbil14}. We adopt the latter interpretation.
This assumption is consistent with the results obtained from the fits.

The moments of inertia of nonlinear creep regions contributing
to the long term response are obtained from the fit parameters
$a_{1} $ and $ a_{2} $. The terms $\alpha_{1}$ and $\alpha_{2}$
yield the numbers of vortices moving inwards and outwards,
respectively, during the glitch. $b$ is the
long term offset of $ \Delta\dot{\Omega}_{c} $ after all the
contributions from creep regions relax back to zero. We interpret
this as the contribution of the change in the external torque. The 
terms with subscripts ``2" and ``3" contribute
$\Delta\dot\Omega_{c}(t)<0$ while the parameter $b$ (external torque
change) and the term with subscript ``1" (inward motion of vortices)
contribute $\Delta\dot\Omega_{c}(t)>0$ (see Figure 1). This term has the longest time constant, $\tau_1\gg \tau_2 > \tau_3$.
The data could also be fitted by assuming no change in the external torque, $b = 0$ and choosing long enough $\tau_1$
so that in the long run $ \Delta\dot{\Omega}_{c} $ relaxes back to zero while accommodating the $ \Delta\dot{\Omega}_{c} (t) > 0$ values for the latest present observations. We have explored models with $ 0 \leq b \leq 1.1 \times 10^{-13}$ rad s$^{-2}$, corresponding to $ -7.2 \times 10^{-4} \leq N_{ext}/N \leq 0$. The values of b between $0.8 \times 10^{-13}$ rad s$^{-2}$ and $1.1 \times 10^{-13}$ rad s$^{-2}$ yield reasonable fit results. Here we choose to explore the possibility of a permanent change in the external torque as reflected by $b = 1 \times 10^{-13}$ rad s$^{-2}$. However, at present we cannot rule out $b = 0$, a full decay. Parameters of the best fits with $b = 0$ and $b = 1 \times 10^{-13}$ rad s$^{-2}$ are shown in Table 1. 
The long term data display quasi-periodic residuals with a period of $\sim 400$ days \citep{antonopoulou14}; we find a best fitting sinusoidal period $P= 394$ d by fitting the data from the last $\sim$1500 days with a model involving only the terms that are dominant in the long term: the contributions from inward moving vortices, the long-term offset with $b = 1 \times 10^{-13}$ rad s$^{-2}$ and the sinusoidal term. In our further investigations comprising all the data we fixed this period for the sinusoid. The residuals of the $b=1 \times 10^{-13}$ rad s$^{-2}$ model also show initial fluctuations, which may be due to transient emission patterns in the magnetosphere. Future timing data will distinguish between these alternatives. 

To apply our extended creep model to the peculiar glitch of PSR J1119-6127,
we use the spin-down rate data for the 2007 glitch, a total of 85 data points. The arrival time data from MJD 54268 indicate that a glitch has taken place since the previous data set on MJD 54220. The first post-glitch data fit to produce frequency derivative values is dated MJD 54300 \citep{weltevrede11}. The presently available spin-down rate and frequency data extending to MJD 56751 was kindly shared with us by Patrick Weltevrede (P. Weltevrede private communication, \cite{antonopoulou14}). The time interval $\Delta$ between the actual glitch date and the first post-glitch frequency derivative values thus lies between $\Delta=32$ days and $\Delta=80$ days. The coefficient $I_{3}/I$ of the exponentially relaxing term is sensitive to the choice of $\Delta$. We arbitrarily take $\Delta=60$ days, which gives $I_{3}/I \cong 1.74 \times 10^{-1}$.

We use the Levendberg-Marquardt method
to find the best fit values of the parameters, starting from initial guesses with
MPFITFUN procedure \citep{markwardt09}\footnote{ http://purl.com/net/mpfit}. The best fit is displayed in Figure 1 and its parameters are listed in Table 1. Inferred model parameter values corresponding to Eq. (15) are shown in Table 2.

\begin{figure}
\centering
\vspace{0.1cm}
\includegraphics[width=1.0\linewidth]{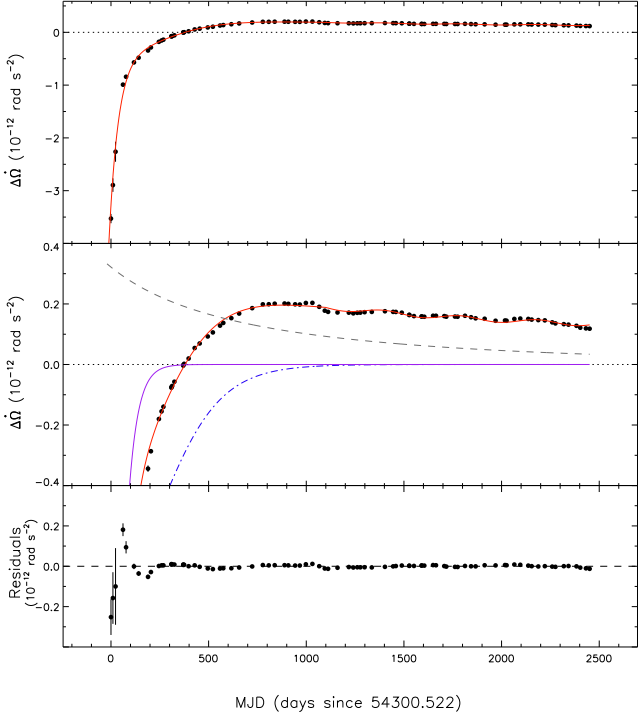}
\caption{Top panel: Fit to the post-glitch spin-down rate data with the model of Eq. (15), with $ \Delta=60 $ days and 
b = 1.0 $\times$ 10$^{-13}$ rad s$^{-2}$. Middle panel: Zoomed version of top panel 
with model components representing contribution of exponential relaxation
term (purple solid line), inward moving vortices (gray dashed line) and outward 
moving vortices (blue dash dotted line) are shown separately. Sinusoidal
component and long-term offset b are not shown in the figure for clarity. 
Bottom panel: Difference between data and model.}
\label{fig:fit}
\end{figure}

\begin{table}
\centering
\caption{Parameters of the best fits to the postglitch frequency derivative data following the 2007 glitch of PSR J1119-6127, with $ \Delta=60 $ days, and $b = 1.0 \times 10^{-13} rad
\;s^{-2} $ (first column) and $b = 0$ (second column) }
\begin{tabular}{lll}
\hline\hline
Parameter & Value (Error) & Value (Error) \\
\hline
$(a_{1})_{-13}(rad\;s^{-2})$  & 1.68 (0.77) & 0.34 (0.29) \\
$(a_{2})_{-13}(rad\;s^{-2})$  & 10.16 (1.43) & 5.93 (0.39)\\
$ (a_{3})_{-12}(rad\;s^{-2}) $ & 9.92 (0.55) & 8.01 (0.38) \\
$ \alpha_{1} $ & -0.68 (0.15) & -0.90 (0.07)\\
$ \alpha_{2} $ & 6.34 (1.48) & 20.65 (3.82)\\
$\tau_{1} (days) $  & 1796 (211) & 20475 (14413)  \\
$\tau_{2} (days) $ & 159 (7) & 129 (4)\\
$\tau_{3} (days) $ & 48 (2) & 58 (2) \\
$(b)_{-13}(rad\;s^{-2})$ & 1.0 & 0.0\\
\hline
\end{tabular}
\end{table}

\begin{table}
\centering
\caption{Inferred Parameters with $b = 1.0\times 10^{-13}rad\;s^{-2}$ .}
\begin{tabular}{ll}
\hline\hline
Parameter & Value \\
\hline
$\left(\frac{I_{A'}}{I}\right) _{-3}$ & 1.11 \\
$\left(\frac{I_{A}}{I}\right) _{-3}$ & 6.70 \\
$\left(\frac{I_{3}}{I}\right) _{-1}$ & 1.74 \\
$ t_{0}'(days) $ & 2046 \\
$ t_{0}(days) $ & 317 \\
$ \left( \delta\Omega'\right)_{-2} (rad\;s^{-1})  $ & 2.69 \\
$ \left( \delta\Omega\right)_{-3} (rad\;s^{-1}) $ & 4.16 \\
$ \left(\frac{\Delta N_{ext}}{N}\right)_{-4} $ & -6.58 \\
\hline
\end{tabular}
\end{table}

The long term remnant $\Delta\Omega_p$ of the glitch in frequency can be found by comparing the indefinite integral of the model for spin-down rate with the observed frequency residual at the latest available data points. Using the frequency residual data on MJD 56688, $\Delta\Omega_p \cong 9.4 \times 10^{-5}$ rad s$^{-1}$ is obtained. We find $\Delta\Omega_p > 0$, unlike the earlier result of Weltevrede et al. (2011), who found a negative long term frequency residual based on the latest post-glitch data then available, but in agreement with their current estimate (Model A in \cite{antonopoulou14}).

\section{Discussion and Conclusions }
\label{sec:discussion}

We have examined the peculiar 2007 glitch of PSR J1119-6127 by
extending the vortex creep model to take into account (i) the possibility
of a glitch associated change in the pulsar external torque, and (ii) inward
motion of vortices. Both of these effects can be induced by a starquake
that triggered the glitch. We model the peculiar glitch of PSR J1119-6127
as follows: a crustquake occurs, causing the crustal plates to move towards
the rotation axis, together with some pinned vortices. At the same time, some
vortices affected by the crustquake are unpinned and move outward.
The glitch is due to the angular momentum transfer associated with the
sudden outward and inward vortex motions. Magnetic field lines, which
move with the conducting crustal plate, change the external torque and give rise to
the abnormal emission properties.

In contrast to the changes in other pulsars' glitches, the long-term 
change in spin-down rate, after transients are over, is (possibly) positive for
PSR J1119-6127. In the creep process under the
action of an external  {\em spin-down} torque, the inward motion of vortices is
thermodynamically unlikely, unless induced by a driving force such as arising
from crustquake induced motion of crustal plates that carrying pinned vortices
inwards. Inward vortex motion increases the lag between local
superfluid and normal matter rotation rates from the steady state value,
thereby accelerating rather than cutting off the creep process. This in turn
increases the rate of angular momentum transfer to the crust and thereby
decreases the spin-down rate of the crust, producing a positive change in the
observed crust spin-down rate.  By contrast, in standard glitches vortices move
outward, decreasing the lag and turning off or suppressing the creep process which
transfers angular momentum from superfluid to the crust; this leads to a negative
step in the spin-down rate of the crust. Glitch induced steps of
either sign arising from the offset in the vortex creep process always relax back to
the pre-glitch spin-down rate as the creep process heals back to the steady state.
The model we have fitted to the spin-down rate data after the 2007 glitch of PSR
J1119-6127 includes the creep response to outward and inward vortex motion as
well as a glitch associated change in the external (pulsar) torque.

\cite{antonopoulou14} fit their data set with two models, model A with a long term exponential relaxation, and model B with a negative frequency second derivative ($\Delta\ddot{\nu}_{p} < 0$ in their notation). The positive $\Delta\dot{\Omega}_{c}$ decays towards zero in both models, asymptotically in the case of Model A. The two models leave comparable residuals. Our investigation of crust breaking, giving a permanent change in the external torque and spin-down rate and causing inward vortex motion, is complementary to their work. Future timing observations will decide if the offset in the spin-down rate is really permanent or relaxing; and searches for thermal signals accompanying future glitches of PSR J1119-6127 will distinguish between the different models.

We must use the long term remnant of the frequency glitch in Eq. (13) to constrain $I_{B}$ and $I_{B'}$, the moments of inertia of superfluid regions which transfer angular momentum {\em only} at glitches, due to the outward and inward motion of unpinned vortices, respectively. Using the values of $\delta\Omega_{s}$ and $\delta\Omega'_{s}$ from Table 2
leads to the constraint $9.4 \times 10^{-5}I= 4.2 \times 10^{-3} I_{B}- 2.7 \times 10^{-2}I_{B'}$. 
This gives $ I_{B}/I > 2.2 \times 10^{-2}$.  Then the superfluid creep regions with a total moment of inertia $I_{s}>I_{A}+I_{B}+I_{A'}+I_{B'}+I_{3}> I_{A}+I_{B}+I_{A'}+I_{3} \gtrsim 20.4\times 10^{-2}I$ is effected by the glitch event. The region with moment of inertia $I_{3} = 1.74 \times 10^{-1}I$ comprises most of the moment of inertia in pinned superfluid. 

Recent calculations \citep{chamel13,andersson12} show that Bragg
scattering of conduction neutrons from nuclei in the
neutron star crust induces a neutron effective mass that is larger than the bare mass. This
``entrainment" of superfluid neutrons in the crust by the crystal
lattice requires that the actual moment of inertia associated with
the superfluid response is larger by a factor $m_n^{*}/ m_n > 1$
where $m_n^{*}$ and $m_n$ are effective and bare neutron masses in
the lattice. The moment of inertia $ I_{s} \gtrsim \; (m_n^{*}/ m_n) \; 20.4 \times
10^{-2}I $ associated with creep cannot be accommodated by the {\em crust} superfluid alone for most neutron star models, 
even without the effective mass correction. In addition to the crust superfluid, other locations are required to sustain vortex creep. Contribution from vortex line-toroidal flux
line pinning and creep at the outer-core of the neutron star
\citep{sidery09}, which has a comparable or larger moment of inertia
than that of the crust superfluid, could provide the required extra
moment of inertia \citep{erbil14}. The moment of inertia of the
creep region where vortex motion is controlled by the toroidal
arrangement of flux lines can amount to $I_{\rm tor}/I \sim 2 \times
10^{-1}$ depending on the radial extension of the toroidal field in
the outer core. Creep here is in the non-linear regime. As no glitch
associated vortex motion is expected, the response to a glitch is
exponential relaxation \citep{erbil14}. The relaxation time
$\tau_{\rm tor} \cong 50$ days for PSR J1119-6117 parameters is in line with
our estimate of $\tau_{3} \cong 48 $ days (and with $\tau_{nl} \cong 33$ days for the 
application to the Vela pulsar, which has parameters similar to those of PSR J1119-6117; \cite{erbil14}). 
So, we argue that $I_{3}/I = 1.74
\times10^{-1}$ reflects the moment of inertia associated with the
toroidal flux line region of the outer core. The moment of inertia
of the {\em crustal} superfluid participating in the glitch, when the
crustal entrainment correction is included, is $I_{s, \rm crust}> (m_n^{*}/
m_n)(I_{A}+I_{B}+I_{A'}) \gtrsim (m_n^{*}/ m_n) \; 2.98 \times
10^{-2}I$. This total moment of inertia fraction can be accommodated in the crust in 
neutron star models with hard equations of state, if the mean value of $ m_n^{*}/ m_n $
to represent the crust superfluid is not much larger than 1. The recent work of \citet{piekarewicz14} shows that the neutron star crust may maintain larger moment of inertia so that the above constraint is easier to be satisfied.

The total numbers of vortices displaced in this glitch are determined by the superfluid
angular velocity changes $\delta\Omega_{s}$ and $\delta\Omega'_{s}$ using  Eq. (9). The number of
vortices that have moved outward is found to be is $\delta N_{out}\sim 1.3\times 10^{13}$, 
while the corresponding number for inward moving vortices is $\delta N_{in}\sim 8.4\times 10^{13}$. 
These numbers are typical of all small or large glitches, from Crab, Vela and other pulsars analyzed so far in terms of vortex unpinning, indicating a particular scale of the glitch trigger.

The glitch associated change in the external torque contributes a constant offset from
the pre-glitch behaviour that remains in the spin-down rate after all post-glitch relaxation
is over. This term, denoted $b$ in Eq. (15), indicates a change in the external torque,
which leads to a change in the spin-down rate through $ \Delta N_{ext}/N_{ext} =
\Delta\dot{\Omega}/\dot{\Omega} + \Delta I /I$. The actual fractional change in the
moment of inertia associated with a possible quake must be less than the
observed glitch magnitude, so $|\Delta I /I| < \Delta\Omega/\Omega \sim
10^{-5} << |\Delta\dot{\Omega}/ \dot{\Omega}| \sim 10^{-4}$. The measured permanent
term $b$ in $ \Delta\dot{\Omega}/ \dot{\Omega}$ therefore gives the fractional change
in the external torque. Taking the external torque to be essentially the dipole radiation
torque, we have:
\begin{equation} 
\frac{\Delta N_{ext}}{N_{ext}}=
3\frac{\Delta\Omega_{c}}{\Omega_{c}}+2\frac{\Delta B_{\bot}}{B_{\bot}} \cong
2\frac{\Delta B_{\bot}}{B_{\bot}},
\end{equation}
as the term $3 \Delta \Omega_{c}/\Omega_{c} \sim 10^{-5}$ is again negligible. We
assume that the magnetic field change is associated with crust breaking, involving broken plates of size $D$ distributed in a ring of the crust of width $D$ and radius $R \cos\alpha$ from the rotation axis, with each plate moving a distance $D$ at the quake. The field
moves with each broken piece of the conducting crust, without any change in the local field
magnitude, and orientation, which we take to be normal to the crust plate. The local field strength varies azimuthally in the broken ring. 
We further assume that the broken ring is in the polar regions of the magnetic field, so that the crust breaking has a strong effect on the external torque. This assumption is plausible if magnetic stresses play a role in crust breaking \citep{franco00, lander14}. A schematic
view of our model for external torque variation via change of magnetic field's perpendicular
component is depicted in Figure 2. The external torque variation is related to the change
$\Delta\alpha$ in the angle between the rotation and magnetic axes:
\begin{equation} \frac{\Delta B_{\bot}}{B_{\bot}}=\frac{\Delta
\alpha}{\tan\alpha}.
\end{equation}
From our estimate of the change in external torque,
$\Delta N_{ext}/N_{ext}\cong -6.58\times 10^{-4} $ given in Table 2, we obtain:
\begin{equation}
\Delta \alpha =
\frac{1}{2} \frac{\Delta N_{ext}}{N_{ext}}\;\tan{\alpha}
\cong (-3.3 \times 10^{-4})\;\tan{\alpha}
\end{equation}
which is between $(-1.0 \times 10^{-4})$ and $(-2.8 \times 10^{-4})$. To obtain this,
we have used the range of $ \alpha $ considered by \citet{weltevrede11},
$\alpha\sim 17\,^{\circ}-30\,^{\circ}$ corresponding to an emission height
$\sim 500$ km and $\alpha\sim 30\,^{\circ}-40\,^{\circ}$ corresponding to an
emission height $\sim$ 1800 km. The tiny change $\Delta\alpha$ in inclination
angle cannot be resolved as an observable glitch associated pulse shape change
in the present radio timing data. Motion of crustal plates towards the pole
($ \Delta \alpha <0 $) results in a reduction in the moment of inertia of the solid,
and therefore an increase in the spin-down rate. This is the signature of a crust
quake in a spinning down pulsar, tending to make the shape more
spherical. The reason for the external torque change is likely to be a starquake, inducing
motion of crustal plates, and reducing $B_{\bot}$, as the surface magnetic field moves
with the conducting plates towards the rotation axis.

The fractional change in moment of inertia due to the motion of the crustal plates is
$\Delta I/I \sim (m/M) \Delta \alpha \lll \Delta \alpha \sim 10^{-4}$, where $m$ is
the total mass of the moving plates and $M$ is the mass of the entire star. Observing the direct
effect of this actual change in the crustal moment of inertia as a glitch is impossible.
The glitch magnitude $ \Delta\Omega $ is due to amplification by the vortex motion triggered 
by crust-breaking and the resulting angular momentum transfer from superfluid to normal matter. We have assumed that some of the vortices pinned to the moving plates are initially carried inward with the plates. This is possible when the increase $\delta \Omega' $ in superfluid rotation rate, due to the inward motion of the pinned vortices on the time scale of crust breaking is not sufficient for the local lag to increase from the steady state value to the critical value, $ \delta \Omega' < \omega_{cr}-\omega_{\infty} $, for typical values of $ \omega_{\infty} $ \citep{alpar89}. 
These vortices will bend and are likely to be strongly perturbed by the
sudden inward motion and become unpinned. The unpinned vortices will then move
downstream azimuthally with the superfluid flow, causing more vortices to unpin and scatter outward until they reach a new radial position where they join the background vortex flow and creep. The avalanche of unpinning takes place rapidly on the glitch ``rise" timescale. Some number $\delta N_{in}$ of vortices associated with the moving plates end up in radial positions inward of their original position while a number $\delta N_{out}$ of vortices end up in radial positions further out compared to their original position. The moment of inertia of the superfluid creep regions affected by the inward motion of the plates and net inward vortex motion is of order
\begin{align}
I_{A'}/I & \cong \frac{4 \pi \rho_{s}  R^{4} D \sin{\alpha} \cos^{2}{\alpha}  }{(2/5)M R^{2}}
\simeq 15/2 \sin{\alpha} \cos^{2}{\alpha}~(D/R) \nonumber\\ & \sim (2D/R),
\end{align}
assuming a uniform density neutron star, and adopting $ \sin{\alpha} \cos^{2}{\alpha} \cong 0.3$ for the range of $\alpha
\cong 17^{\circ}-40^{\circ}$ indicated by \cite{weltevrede11}. Using the value
of $ I_{A'}/I $ from fits, we obtain the ring width $ D \sim  6R_{6} $ m, where
$R_{6}$ is the neutron star radius in units of $10^{6}$ cm. The number of vortices
pinned to each plate is
\begin{equation}
 \delta N_{plate} \sim D^{2} \dfrac{2\Omega}{\kappa}
\sim 3\times10^{8}\Omega R_{6}^{2} \sim 5.5 R_{6}^{2}\times 10^{9}.
\end{equation}
The total number of vortices associated with broken plates with a net inward motion during the glitch, is $ \delta N_{in} \sim 8.4\times
10^{13}$ vortices, as we estimated above from the results of our fits for $ t_{0}'$, a parameter independent from $I_{A'}/I$ which we used to estimate the plate size $D$. The number of plates involved should be $\sim \delta N_{in} / \delta N_{plate} \sim 10^{4}$, in agreement with the number of plates in the broken ring,  $\sim 2\pi R/D \sim 10^{4}$. A comparable
number of vortices $ \delta N_{out} \sim 1.3\times 10^{13}$ end up moving outward
through a superfluid region of comparable moment of inertia, $ I_{A} $. We find here an
indication that the common scale, $ \sim 10^{13} $, of the number of vortices unpinned
in all pulsar glitches may be associated with the number of vortices in the typical plate
size $D$ involved in a triggering crust quake, multiplied by the number of plates involved, $\sim 2\pi R/D \sim 10^{4}$.
These scales rest on the single parameter, the plate size $D$ which must be related to the physics
of the crustal solid. This plate size $D$ is of the same order of magnitude as the `mountain' 
height $\sim 1$ m estimated for the Crab pulsar \citep{chamel08}. Note that the critical strain angle $ \theta_{cr}$ at which the crust lattice breaks is $ \theta_{cr} \sim D/h $ where $h$ is the radial thickness of the broken crustal plates. Thus,
\begin{equation}
\theta_{cr} \sim 10^{-2}\left(\dfrac{D}{1 \; m} \right) \left( \dfrac{h}{100 \; m}\right)^{-1}, \nonumber 
\end{equation}
compatible with the results of \citet{horowitz09} for the critical strain angle.

It is interesting to compare the moment of inertia fractions in {\em crust} superfluid regions through
which the unpinned vortices moved during the peculiar glitch of PSR J1119-6127, given in Table 2, with 
the corresponding Crab \citep{alpar96} and Vela \citep{alpar93} values,
$(0.01-1.87)\times 10^{-3}$ and $(2.3-3.4)\times 10^{-2}$ respectively, as this gives a lower limit on the moment of inertia fraction in the crust, leading to constraints on the neutron star equation of state \citep{datta93,link99}.
With its characteristic age of  $ \sim 1625 $ years, PSR J1119-6127 is between the Crab and Vela pulsars in age, but its implied crustal superfluid  moment of inertia fraction $2.98\times 10^{-2}$ is comparable to the values inferred for the Vela pulsar. 
The qualitative evolution of glitching behaviour from Crab-like to Vela-like was proposed
to be due to the development of connections in a network of vortex creep regions, so
that the moment of inertia involved increases with age \citep{pines85}. PSR J1119-6127
should already have a sufficiently well connected vortex creep network. Presumably the
high magnetic field and associated stresses in the crust of this pulsar lead to high crust
breaking activity. While similar moments of inertia in vortex creep regions $I_A$ are inferred, and the long term fractional offsets in the spin-down rate are similar in absolute value, for the
Crab case no change in electromagnetic signature is observed.

Magnetic stresses will play a role comparable to that of rotation induced stresses in conventional starquake models if, roughly, 
\begin{equation}
\dfrac{B_{0}^{2}-B^{2}}{8\pi} \sim \dfrac{1}{2}\rho R^{2}\left( \Omega_{0}^{2}-\Omega^{2}\right) \sim \dfrac{1}{2}\rho R^{2} \Omega \mid \dot{\Omega} \mid t_{g}, \nonumber
\end{equation}
where $ B_{0} $ and $ \Omega_{0} $ denote reference values of $ B $ and $ \Omega $ frozen into the crust. Using the glitch interval $ t_{g} \sim 3 $ yrs, typical for young pulsars, we find that magnetic stresses can play a role where $ B \gtrsim 10^{13} $ G in the crust. The magnetic field can have poloidal and toroidal components whose geometry will determine where in the crust the local stresses reach the critical values for crust breaking \citep{lander14}. The higher multiple components of the magnetic field and the geometry of the stress tensor when both magnetic and rotational effects are included further complicate the situation. In addition, the changes in electromagnetic signature, as seen only in PSR J1119-6127, are likely to occur if the broken plate extends to the surface and leads to reconfiguration of the magnetosphere. This may explain why the behaviour exhibited by PSR J1119-6127 is rare. As a rough guideline, such behaviour may be exhibited by young pulsars with high magnetic field in young pulsars with high magnetic field. 

We should note the different responses of the crust and the superfluid to a starquake.
After the crust breaks and plates move towards the rotation axis in order to relieve their stresses,
those broken pieces of the crust are stuck to new metastable positions, and do not come back to their pre-glitch sites. Thus, crust breaking and crustal motion
are irreversible. Starquake induced inward motion of vortices leads to a local excess
of vortex lines and thereby to faster rotation of the superfluid. Creep becomes more
efficient, evolving back towards steady state with an enhanced vortex current, as
described by the vortex creep response employed in our fits. In the end, the superfluid
relaxes back to the pre-glitch dynamical steady state, and all the parts of
crustal superfluid and the rest of the star spin-down at the same rate. Indeed, without structural changes,
the response of all internal torques will always relax back to $ \Delta\dot{\Omega}_c = 0$.

The main reason of the switch in emission patterns to intermittent and RRAT
behaviour lasting for about a hundred days after the event in 2007 is likely to
be the effect of the quake on magnetic field lines which are anchored to the crust.
If the magnetospheric field pattern could move rigidly, without distortion,
together with the motion of the crustal plate, as a consequence of the shift by $\Delta\alpha \sim
10^{-4}$, there would be no significant change in the emission pattern. However,
when a crustal plate moves in a quake, the elastic response of the
field lines, which twist and reconnect, can amplify a small shift in the crustal
position into a complex and drastic change in the emission pattern, helped by
rotation which results in twisting of magnetic field lines anchored to the highly
conducting crust \citep{beloborodov09}.  The distorted magnetospheric
geometry will subsequently relax towards a quasi-stable configuration for the
new position of the plate. The changes in the emission pattern are observed
for about a hundred days following the glitch, after which the pulsar returns to its
pre-glitch pulse shape and emission pattern. Twisting of field lines and their
subsequent relaxation will also introduce temporary fluctuations in arrival times.
Our timing model fits indeed leave relatively large residuals for about a hundred
days in the post-glitch data given in the bottom panel of Figure 1. 

\begin{figure}
\centering
\includegraphics*[width=1.0\linewidth]{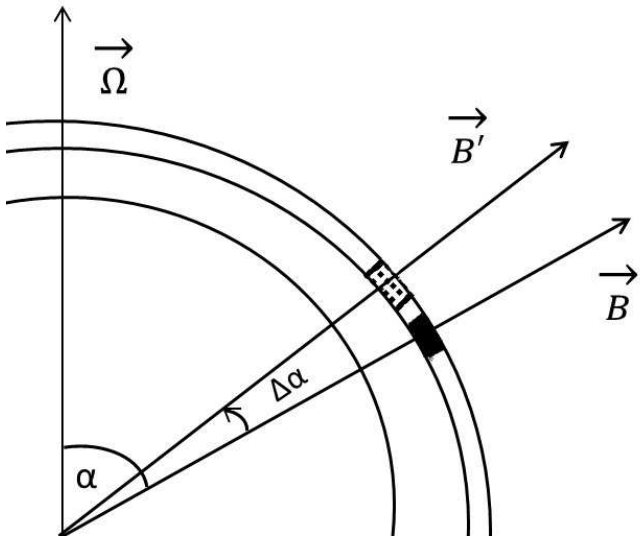}
\caption{Starquake Model in cross section. The dotted area represents new position of crustal plates (broken ring) after the starquake.}
\label{fig:sekil2}
\end{figure}

Although the recently discovered magnetar anti-glich from 1E 2259+586 by
\citet{archibald13} has also shown a negative spin jump and changing emission
features, the situation is very different from that of PSR J1119-6127. In  the case of
1E 2259+586, the magnitude of the change in spin-down rate, $|\Delta \dot \Omega|
\sim 2.8|\dot \Omega|$ is too large to be associated with the superfluid regions
in the star. It is likely that only a large change in the external torque is 
involved, as suggested by the violent change in emission, so that this magnetar `antiglitch' 
must be external/magnetospheric in origin \citep{lyutikov13, tong14}.

In summary, the peculiar glitch of PSR J1119-6127 offers an invaluable opportunity
for the reexamination and extension of glitch models to account for anomalous
glitch signatures and transient emission phenomena initiated by a quake leading
to a change in the external torque and triggering the response of the superfluid
regions of the neutron star. Our model predicts that the change in the external torque is permanent. The coincidence of the numbers of vortices involved in the glitch with the numbers inferred in Crab and Vela pulsar glitches is highly suggestive, supporting the explanation in terms of a crust breaking event with a typical plate size, which may be a common, even universal, trigger for glitches. If future timing observations rule out a permanent change in the external torque, this coincidence would turn out to be spurious. The explanation of the post-glitch evolution of $\Delta\dot\Omega_{c}(t)$ in terms of internal torques responding to glitch associated inward and outward vortex motion, and relaxing to  $\Delta\dot\Omega_{c}=0$ is viable when there are no observed changes in the external torque. 

\section*{Acknowledgments}
This work is supported by the Scientific and Technological Research Council of Turkey
(T\"{U}B\.{I}TAK) under the grant 113F354. M.A.A. is a member of the Science Academy
(Bilim Akademisi), Turkey. We thank Patrick Weltevrede for kindly providing us with
unpublished post-glitch timing data, and Danai Antonopoulou, Anna Watts,
Cristobal Espinoza and Patrick Weltevrede for useful discussions and for sharing with
us their preprint with an alternative approach to the post-glitch behaviour of PSR J1119-6127. We thank the referee for useful comments.


\end{document}